\documentstyle[12pt]{article}
\def\ave#1{\left\langle #1\right\rangle}
\def\half{{\textstyle\frac{1}{2}}}

\font\rm=cmr10 scaled\magstep1
\def\re{\hbox{\rm Re}}
\def\im{\hbox{\rm Im}}
\def\per{\hbox{\rm per}\,}
\def\sper{\hbox{\rm sper}\,}
\def\diag{\hbox{\rm diag}\,}

\newcommand{\sgn}[1]{\hbox{\rm sign}(#1)}

\newcommand{\ve}[1]{\hbox{\boldmath{$#1$}}}

\newcommand{\ma}[1]{\hbox{\boldmath{\rm #1}}}
\newcommand{\ti}[1]{\tilde{#1}}

\begin{document}
\begin{flushright}
\today
\end{flushright}
\begin{center}
\vspace{0.3in}
{\Large\bf  Parametric statistics of zeros of Husimi representations of 
quantum chaotic eigenstates and random polynomials }\\
\vspace{0.4in}
\large
Toma\v z Prosen
\footnote{e-mail: prosen@fiz.uni-lj.si}\\
\normalsize
\vspace{0.3in}
Physics Department, Faculty of Mathematics and Physics,\\
University of Ljubljana, Jadranska 19, 1000 Ljubljana, Slovenia\\
\vspace{0.3in}
\end{center}
\vspace{0.5in}

\noindent{\bf Abstract}
Local parametric statistics of zeros of Husimi representations of quantum
eigenstates are introduced. It is conjectured that for a classically fully 
chaotic systems one should use the model of parametric statistics of complex
roots of Gaussian random polynomials which is exactly solvable as demonstrated 
below. For example, the velocities (derivatives of zeros of Husimi function 
with respect to an external parameter) are predicted to obey a universal 
(non-Maxwellian) distribution
$$\frac{d{\cal P}(v)}{dv^2} = \frac{2}{\pi\sigma^2}(1 + |v|^2/\sigma^2)^{-3},$$ 
where $\sigma^2$ is the mean square velocity.
The conjecture is demonstrated numerically in 
a generic chaotic system with two degrees of freedom. Dynamical formulation 
of the ``zero--flow'' in terms of an integrable many-body dynamical system 
is given as well. 
\vspace{0.7in}

\noindent
\section{Introduction}
The intense research in the so-called quantum chaology has produced many
different signatures of classical chaos in the corresponding quantum 
Hamiltonian systems. It has been found that energy spectra and eigenstates of 
classically fully chaotic quantum systems have universal statistical properties
which can be described by stochastic models with no free parameters, like 
random matrix theory (RMT). 

In the Bargmann (or Husimi) representation, eigenstates 
of quantum systems can be uniquely represented in terms of complex analytic 
functions in phase space variables. Quite recently, it has been proposed 
\cite{LV93} that in case of 1-dim systems (or Poincar\'e surface of section 
reductions of 2-dim systems \cite{P95}), where phase space is two dimensional, 
and one has only one complex variable $z=q + ip$, any eigenstate can be
uniquely represented by the collection of (complex) zeros of its Bargmann or
Husimi representation. This has been called the stellar representation. It has
been found \cite{P95} that the structure of zeros of Husimi representation of 
an eigenstate is reminiscent of the structure of classical phase space:
(i) in classically fully chaotic systems the zeros tend to spread uniformly
over the whole classically allowed region of phase space \cite{LV93},
(ii) in classically integrable systems the zeros lie on 1-dim.
anti-Stokes curves \cite{LV93}, (iii) while in generic mixed systems a part
of zeros uniformly cover chaotic components of classical phase space, zeros
in regular islands lie along 1-dim. classically invariant curves - tori, while 
substantial part of zeros lie on 1-dim. classically noninvariant
(generalized anti-Stokes) curves. 

In the following we shall be interested only in the case of
classically fully chaotic systems. Let us assume that the Hamiltonian of
our chaotic system can be statistically described by Gaussian 
orthogonal/unitary (in the presence/absence of anti-unitary symmetry) 
random matrix in a generic basis. If we choose the harmonic basis then the 
Bargmann representation of an arbitrary eigenstate is given by the entire 
function 
\begin{equation}
f(z) = \sum_{n=0}^{\infty} \frac{c_n}{\sqrt{n!}} z^n,
\label{eq:1}
\end{equation}
and the Husimi representation in appropriately scaled units is 
$|f(z)|^2 e^{-|z|^2}$, where $z=q+ip$.
$c_n$ are the coefficients of the expansion of that eigenstate in the
harmonic basis, which are, by assumption and hence according to RMT, 
real/complex Gaussian (pseudo) random variables. 
Ideally, as RMT predicts, there 
should be no correlations among them $\ave{c_n c^*_m} = \delta_{nm}$, 
although we believe 
that for nonzero shortrange correlations among the coefficients $c_n$ our 
general conclusions are still valid. Since Taylor expansion of $f(z)$
around arbitrary point $z_0$ is convergent one may define and study
{\em Gaussian random polynomials} of order $N$ and the statistical properties 
of their roots and in the end, if neccesary, take the `thermodynamic' limit
$N\rightarrow\infty$ (in a sense of increasing number of zeros --- 
quasi-particles) of Gaussian random holomorphic functions (\ref{eq:1}).
The limit $N\rightarrow\infty$ is in fact compatible with a semiclassical 
limit $\hbar\rightarrow 0$ since the order $N$ of a Taylor polynomial
aproximating a Bargmann function (and its zeros) should be at least equal to 
the number of basis states covering the classically accessible area $A$ of 
phase space (or surface of section), $N > A/(2\pi\hbar)$, while higher 
coefficients $a_{N+n},\,n > 0$, vanish rapidly with increasing $n$.
One can also consider cases of different geometry, for example of kicked 
spin $j$ systems, where quantum (eigen)states can be exactly represented in 
terms of the so-called SU(2) polynomials of finite order $2j$.

It has been shown recently by Hannay
\cite{H96} for the case of complex coefficients and supplemented by Prosen
\cite{P96} for the case of real coefficients that the statistics 
of zeros of Gaussian random polynomials are exactly solvable and all
k-point correlation functions can be given in terms of simple analytical
formulae. It has been demonstrated \cite{LS95,P95u} that 
the results obtained by random polynomials indeed reproduce the statistics of 
zeros of Husimi representations of chaotic eigenstates.  
\\\\
It is the aim of this paper to introduce the {\em local} parametric statistics 
of zeros of Husimi representations of the quantum eigenstates. 
Let us take a family of Hamiltonian systems which smoothly depend upon
an external parameter $\lambda$. Then the Bargmann representation of
a given eigenstate $f(z,\lambda)$ and its zeros
$z_n(\lambda)$ are also smooth functions of the parameter $\lambda$.
Therefore we can introduce the {\em velocities} $v_n$ as derivatives of zeros 
$z_n$ with respect to an external parameter $\lambda$
\begin{equation}
v_n(\lambda) = \frac{d}{d\lambda}z_n(\lambda) = 
-\frac{\partial_{\lambda}f(z_n,\lambda)}{\partial_z f(z_n,\lambda)}
\end{equation}
Then we define a parametric k-point correlation function of $k$ zeros
$\ve{z}=(z_1,\ldots,z_k)$ and $k$ velocities $\ve{v}=(v_1,\ldots,v_k)$ as
\begin{equation}
\ti{\rho}_k(\ve{z},\ve{v}) = 
\ave{\delta^{2k}(\ve{z}-\ve{z}^\prime)
\delta^{2k}(\ve{v}-\ve{v}^\prime)}_{\ve{z}^\prime,\ve{v}^\prime}
\label{eq:pcf}
\end{equation}
where $\ave{}_{\ve{z}^\prime,\ve{v}^\prime}$ represents an (ensemble) average
over all k-tuples of zeros $\ve{z}^\prime$ and corresponding velocities 
$\ve{v}^\prime$ of a given state (or ensemble of states). 
In other words, $\ti{\rho}_k(\ve{z},\ve{v})d^{2k}\ve{z}\, d^{2k}\ve{v}$
is a probability to find a k-tuple of zeros and corresponding velocities
in a small $4k$-cube of volume $d^{2k}\ve{z}\, d^{2k}\ve{v}$ around point 
$(\ve{z},\ve{v})$. Integrating out the velocities one should obtain the
usual $k$-point correlation function $\rho_k(\ve{z})$ of zeros only
\cite{H96,P96}
\begin{equation}
\rho(\ve{z}) = \int d^{2k}\ve{v} \ti{\rho}(\ve{z},\ve{v}).
\label{eq:rhotirho}
\end{equation}
We show below, in section 2, that this parametric
correlation functions can be explicitly calculated for Gaussian ensembles
of parameter-dependent random polynomials with either complex or real 
coefficients. In section 3, we shall verify our conjecture that the
obtained result on parametric statistics of roots of statistical ensembles of
random polynomials 
may be applied to quantum chaotic systems by presenting some
numerical results obtained in a generic chaotic system, namely 2-dim
semi-separable oscillator \cite{P95a,P95}. 
In section 4, we write a closed system of `equations of motion'
for the zero-flow $z_n(\lambda)$ for the simplest, linear parametric 
dependence of the coefficients $a_n$, and stress the integrability of the
underlying dynamical system. 

\section{Parametric statistics of roots of Gaussian random polynomials}

\subsection{Complex coefficients}

In this section we study parametric statistics of statistical ensembles of
Gaussian random polynomials. In the first subsection we are dealing with the
case of complex coefficients while slightly more complicated case of
real coefficients will be dealt with in the next subsection.
Let us write a random polynomial of order $N$ in a form
\begin{equation}
f(z,\lambda) = \sum_{n=0}^N a_n(\lambda)z^n
\end{equation}
where the coefficients $a_n(\lambda)$ (for a fixed realization) may depend
smoothly on an external parameter $\lambda$. 
We will need only first derivatives with respect to $\lambda$ so the
parametric statistical Gaussian ensemble of random polynomials is
completely specified by fixing $\lambda$ and saying that 
$a_n$ and $\partial_\lambda a_n$ are complex Gaussian random variables with
prescribed covariances 
$$\ave{a_n a^*_m},\quad \ave{a_n \partial_\lambda a^*_m},\quad
\ave{\partial_\lambda a_n \partial_\lambda a^*_m}$$
which need not be further specified for the statement of the general result.
Fixing the two $k$-tuples of complex numbers $\ve{z}$ and $\ve{v}$, we define
the $3k$ linear combinations of $2(N+1)$ Gaussian random variables
$a_n,\partial_\lambda a_n, n=0,\ldots,N$
\begin{eqnarray}
f_j &=& f(z_j,\lambda) = \sum_{n=0}^N a_n z_j^n,\label{eq:deff}\\
f_j^\prime &=& \partial_z f(z_j,\lambda)=\sum_{n=0}^{N-1} (n+1)a_{n+1} z_j^n,
\label{eq:defdf}\\
\ti{f}_j &=& \frac{d}{d\lambda}f(z_j,\lambda) = \bar{f}_j + v_j f^\prime_j,
\quad \bar{f}_j = \partial_\lambda f(z_j,\lambda)
=\sum_{n=0}^N \partial_\lambda a_n z_j^n \label{eq:deftif}
\end{eqnarray}
which are again Gaussian random variables. The joint probability
distribution of $3k$ random variables $\ve{\xi}=(\ve{f},\ve{f}^\prime,
\ti{\ve{f}})$ can be therefore written as
\begin{equation}
P(\ve{f},\ve{f}^\prime,\ti{\ve{f}}) = \frac{1}{\pi^{3k}\det\ti{\ma{M}}}
\exp\left(-\ve{\xi}^*\cdot\ti{\ma{M}}^{-1}\ve{\xi}\right)
\end{equation}
where $\ti{\ma{M}} = \ti{\ma{M}}(\ve{z},\ve{v})$ is 
$3k\times 3k$ Hermitian covariance matrix which is written in a block form as
\begin{eqnarray}
\ti{\ma{M}} &=& \ave{\ve{\xi} \otimes \ve{\xi}^*} = 
  \pmatrix{\ma{A} & \ma{B} & \ti{\ma{D}}\cr
  \ma{B}^\dagger & \ma{C} & \ti{\ma{E}}\cr
  \ti{\ma{D}}^\dagger & \ti{\ma{E}}^\dagger & \ti{\ma{F}}\cr} \label{eq:defM}\\
A_{jl} &=& \ave{f_j f^*_l},\quad
B_{jl} = \ave{f_j {f_l^\prime}^*},\quad
C_{jl} = \ave{f^\prime_j {f_l^\prime}^*},\label{eq:defABC}\\
\ti{D}_{jl} &=& \ave{f_j \ti{f}^*_l},\quad
\ti{E}_{jl} = \ave{f^\prime_j \ti{f}^*_l},\quad
\ti{F}_{jl} = \ave{\ti{f}_j \ti{f}^*_l}.\label{eq:defDEF}
\end{eqnarray}
Note that tilded symbols are used to denote matrices (or vectors or scalars) 
which explicitly depend on the parametric velocities $\ve{v}$.
$k$-tuple $\ve{z}$ are the zeros if $\ve{f}=0$, and $k$-tuple $\ve{v}$ are the 
velocities if in addition $\ti{\ve{f}} \equiv (d/d\lambda)\ve{f} = 0$. 
So the parametric $k$-point correlation function (\ref{eq:pcf}) can be written 
as a linear transformation of a joint distribution $P(\ve{\xi})$
\begin{eqnarray}
\ti{\rho}(\ve{z},\ve{v}) &=& \int d^{2k} \ve{f}^\prime 
\frac{\partial (\ve{f},\ti{\ve{f}})}{\partial (\ve{z},\ve{v})}
P(\ve{0},\ve{f}^\prime,\ve{0})\\
&=& \frac{1}{\pi^{3 k}\det\ma{M}}\int
\prod_{j=1}^k d^2 f^\prime_j\,|f^\prime_j|^4
\exp\left(-{\ve{f}^\prime}^*\cdot\ti{\ma{L}}^{-1}\ve{f}^\prime\right)
\label{eq:Gauss}
\end{eqnarray}
where
$\partial (\ve{f},\ti{\ve{f}})/\partial (\ve{z},\ve{v})
=\prod_{j=1}^k |f^\prime_j|^4$ is the Jacobian of the mapping 
$(\ve{z},\ve{v})\rightarrow (\ve{f},\ti{\ve{f}})$, and
\begin{equation}
\ti{\ma{L}} = \ma{C}-\ma{B}^\dagger\ma{A}^{-1}\ma{B}-
(\ti{\ma{E}}-\ma{B}^\dagger\ma{A}^{-1}\ti{\ma{D}})
(\ti{\ma{F}}-\ti{\ma{D}}^\dagger\ma{A}^{-1}\ti{\ma{D}})^{-1}
(\ti{\ma{E}}^\dagger-\ti{\ma{D}}^\dagger\ma{A}^{-1}\ma{B})
\label{eq:defL0}
\end{equation}
is the central $k\times k$ block of the inverse of covariance matrix, 
$\ti{\ma{M}}^{-1}$.
The dependence on positions of zeros $\ve{z}$ and velocities $\ve{v}$ is 
digged in the definitions of the matrices 
(\ref{eq:defM},\ref{eq:defABC},\ref{eq:defDEF}) In general, the 
dependence on velocities can be made explicit in the following way. 
Writing a diagonal velocity matrix as $\ma{V}=\diag\{v_j,j=1\ldots k\}$ and 
using a definition (\ref{eq:deftif}) one can observe that the covariance 
matrices $\ti{\ma{D}},\ti{\ma{E}},\ti{\ma{F}}$ have a simple velocity 
dependence which can be written in terms of their velocity-independent 
counterparts $\ma{D},\ma{E},\ma{F}$
\begin{eqnarray}
\ti{\ma{D}} &=& \ma{D} + \ma{B}\ma{V}^\dagger,\quad D_{jl} = 
\ave{f_j \bar{f}^*_l},\nonumber \\
\ti{\ma{E}} &=& \ma{E} + \ma{C}\ma{V}^\dagger,\quad E_{jl} =
\ave{f^\prime_j \bar{f}^*_l}, \label{eq:defDEF2}\\
\ti{\ma{F}} &=& \ma{F} + \ma{E}^\dagger\ma{V}^\dagger + \ma{V}\ma{E} +
\ma{V}\ma{C}\ma{V}^\dagger,\quad F_{jl} = \ave{\bar{f}_j \bar{f}^*_l}\nonumber
\end{eqnarray}
This relations can be used to prove that the determinant of the covariance
matrix does not depend on velocities
\begin{equation}
\det\ti{\ma{M}} = \det\ma{M}
\end{equation}
where $3k\times 3k$ matrix $\ma{M}$ is obtained from $\ti{\ma{M}}$ by replacing
the blocks $\ti{\ma{D}},\ti{\ma{E}},\ti{\ma{F}}$ by $\ma{D},\ma{E},\ma{F}$.
Using some elementary algebra one can rewrite the matrix $\ti{\ma{L}}$ in the
form which makes velocity dependence explicit
\begin{equation}
\ti{\ma{L}}^{-1} = \ma{G}^{-1} + (\ma{V}^\dagger + \ma{G}^{-1}\ma{K})
(\ma{H}-\ma{K}^\dagger\ma{G}^{-1}\ma{K})^{-1}
(\ma{V} + \ma{K}^\dagger\ma{G}^{-1})
\label{eq:defL}
\end{equation}
where we have introduced the matrices
\begin{eqnarray}
\ma{G} &=& \ma{C}-\ma{B}^\dagger\ma{A}^{-1}\ma{B} = \ma{G}^\dagger,\nonumber\\
\ma{H} &=& \ma{F}-\ma{D}^\dagger\ma{A}^{-1}\ma{D} = \ma{H}^\dagger,
\label{eq:defGHK}\\
\ma{K} &=& \ma{E}-\ma{B}^\dagger\ma{A}^{-1}\ma{D}.\nonumber
\end{eqnarray}
The representation of parametric correlations $\ti{\rho}_k$ in terms of 
moments of a Gaussian (\ref{eq:Gauss}) is very convenient
since it may be explicitly evaluated as the sum of all possible pairwise
contractions of integration variables $f^\prime_j$ (Wick theorem) and 
expressed in a compact form following an approach of Hannay \cite{H96}
\begin{equation}
\ti{\rho}_k(\ve{z},\ve{v}) = \frac{\det\ti{\ma{L}}}{\pi^{2k}\det\ma{M}}
\per\pmatrix{\ti{\ma{L}} & \ti{\ma{L}}\cr \ti{\ma{L}} & \ti{\ma{L}}\cr}
\label{eq:parstc}
\end{equation}
where {\em permanent} of a square matrix
$\per{S} = \sum_{p}\prod_j S_{j p_j}$ is a symmetric analog of a 
determinant $\det{S} = \sum_{p} (-)^p \prod_j S_{j p_j}$ where $p$
are permutations with signatures $(-)^p$. Integrating out the velocities, 
which can be done by putting expression (\ref{eq:defL}) into eq. 
(\ref{eq:Gauss}) and evaluating the inner Gaussian integrals in terms of
new variables $u_j = v_j f^\prime_j$,
one obtains the $k-$point correlation function of Hannay \cite{H96}
\begin{equation}
\rho_k(\ve{z}) = \frac{\per{\ma{G}}}{\pi^k \det{\ma{A}}}.
\end{equation}
The formula (\ref{eq:parstc}) is a general result on parametric statistics
of Gaussian random polynomials with complex coefficients. Its important
feature is purely algebraic dependence on velocities in contradistinction
with e.g. parametric energy level statistics (see e.g. \cite{H91}, chapter 6)
where velocities have Maxwellian distribution since the Hamiltonian of the 
energy level flow can be clearly written as the sum of the usual kinetic and 
potential part.

In the important special case where (the
coefficients of) the random polynomial and its parametric derivative are
statistically uncorrelated
\begin{equation}
\ave{a_n \partial_\lambda a_m}=0
\label{eq:driftless}
\end{equation}
we obtain parametric correlation functions which are invariant under
the change of sign of velocities 
$\ti{\rho}_k(\ve{z},-\ve{v})=\ti{\rho}_k(\ve{z},\ve{v})$ 
since $\ma{E}=\ma{D}=\ma{K}=0$ and therefore
$$\ti{\ma{L}}^{-1} = \ma{G}^{-1} + \ma{V}^\dagger \ma{F}^{-1}\ma{V}.$$ In other
words, the average velocity (and all its odd moments) is zero $\ave{\ve{v}}=0$.
In this case, the determinant of $3k\times 3k$ covariance matrix has also
a simple factorization in terms of $k\times k$ matrices
$$\det\ma{M} = \det\ma{A}\det\ma{G}\det\ma{F}.$$

Since the $2-$point parametric correlation function depends on
$4$-complex arguments it may be useful 
to define also the 2-point velocity-moments of the parametric correlation 
functions
\begin{equation}
\ave{v_1^k v_2^l {v^*_1}^m {v^*_2}^n} = 
\frac{1}{\rho_2(z_1,z_2)}\int d^2 v_1 d^2 v_2\, v_1^k v_2^l {v^*_1}^m {v^*_2}^n
\ti{\rho}_2(z_1,z_2,v_1,v_2)
\label{eq:velmom}
\end{equation}
which are different from zero only if $k+l=m+n$ and finite if
$k+l+m+n\le 4$. The nontrivial velocity-moments, which still depend on the
positions of two zeros $z_1$ and $z_2$, can be calculated using the Wick
theorem from the representation (\ref{eq:Gauss}). Let us quote the results
for the {\em driftless} case (\ref{eq:driftless})
\begin{eqnarray}
\ave{v_1 v_2^*} &=& G_{21} F_{12}/(G_{11}G_{22} + G_{12}G_{21}),\\
\ave{|v_1|^2 |v_2|^2} &=& (F_{11}F_{22} + F_{12}F_{21})/
(G_{11}G_{22} + G_{12}G_{21}),\\
\ave{v_1^2 {v^*_2}^2} &=& 2 F_{12}^2/(G_{11}G_{22} + G_{12}G_{21}).
\end{eqnarray}

So far the correlations between the coefficients of random polynomials
$a_n,\partial_\lambda a_n$ have been completely arbitrary! 
Now we specialize to the case where coefficients of random polynomials
are $\delta-$correlated 
\begin{equation}
\ave{a_n a^*_m} = \delta_{mn} b_n,\quad
\ave{\partial_\lambda a_n \partial_\lambda a^*_m} = \delta_{mn} \bar{b}_n
\end{equation}
where the variances $b_n,\bar{b}_n$ are still arbitrary.
Introducing two polynomials
\begin{equation}
g(s) = \sum_{n=0}^N b_n s^n,\quad \bar{g}(s) = \sum_{n=0}^N \bar{b}_n s^n
\end{equation}
the relevant matrices can be expressed as
\begin{eqnarray}
A_{jl}(\ve{z}) &=& g(z_j z^*_l),\\
B_{jl}(\ve{z}) &=& z_j g^\prime (z_j z^*_l),\\
C_{jl}(\ve{z}) &=& 
g^\prime (z_j z^*_l) + z_j z^*_l g^{\prime\prime}(z_j z^*_l),\\
F_{jl}(\ve{z}) &=& \bar{g}(z_j z^*_l).
\end{eqnarray}
Putting $k=1$, $1-$point parametric statistics can be explicitly written in an 
elegant factorized form
\begin{equation}
\ti{\rho}_1(z,v) = \rho_1(z)\frac{\ti{\nu}(v/\sigma(z))}{\sigma^2(z)}
\end{equation}
where
\begin{equation}
\rho_1(z) = \frac{1}{\pi}\frac{d}{ds}s\frac{d}{ds}\log g(s)
\Big\vert_{s=|z|^2}
\end{equation}
is a general density of zeros as can be deduced from \cite{H96} and 
\begin{equation}
\ti{\nu}(v) = \frac{2}{\pi}(1 + |v|^2)^{-3}
\label{eq:veldis}
\end{equation}
is an universal velocity distribution normalized to a unit mean square and
$\sigma^2(z) = \ave{|v|^2}$ is a mean square velocity which is inversely 
proportional to the density of zeros
\begin{equation}
\sigma^2(z) = \frac{\bar{g}(|z|^2)}{g(|z|^2)}\frac{1}{\pi\rho_1(z)}.
\end{equation}
So, the theory of random polynomials predicts a universal form of a
velocity distribution (\ref{eq:veldis}) when it is locally rescaled to a unit 
mean square local velocity. 

In case of eigenstates of RMT in the Bargmann
representation one has $b_n = 1/n!, \bar{b}_n = \sigma^2/n!$
and $N\rightarrow\infty$, so 
\begin{equation}
g(s) =\exp(s),\quad \bar{g}(s) = \sigma^2\exp(s)
\end{equation}
and therefore the density distribution and the local mean
square velocity are constant, $\rho_1(z) = 1/\pi, \ave{|v|^2} = \sigma^2$,
so one has
\begin{equation}
\ti{\rho}_1(z,v) = \frac{2}{\pi^2\sigma^2}(1 + |v|^2/\sigma^2)^{-3} 
\end{equation}
In this probably the most important particular case we are also able to give 
some details of 2-point parametric correlation function 
$\ti{\rho}_2(z_1,z_2,v_1,v_2)$ which is only a function of the 4 real
quantities instead of 8: 
distance between roots $|z_2-z_1|$, magnitudes of the velocities
$u_1=|v_1|,u_2=|v_2|$ and the angle between velocities 
$\phi=\arccos(\re v_1 v_2^*/|v_1 v_2|)$. Writing $s = |z_2-z_1|^2$ and fixing 
the velocity scale by putting $\sigma=1$ we may express $2-$point 
parametric correlations as
\begin{eqnarray}
\ti{\rho}_2 &=& \frac{4(e^s - 1)^5\alpha^5 (\beta^2 + \gamma^2 + 4\beta\gamma)}
{\pi^2 (\beta - \gamma)^5},\\
\alpha &=& e^{2s} - (s^2+2) e^s + 1 \nonumber \\
\beta &=& e^s ((e^s-1)(e^s-1-s) + \alpha u_1^2)
((e^s-1)(e^s-1-s) + \alpha u_2^2)
\nonumber \\
\gamma &=& e^{2s}(e^s-1)^2(e^{-s}-1+s)^2 + \alpha^2 u_1^2 u_2^2 
- 2 \alpha e^s (e^s - 1)(e^{-s}-1+t) u_1 u_2 \cos(\phi) \nonumber 
\end{eqnarray}
Since this expression is quite complicated it is worthwhile to study its
asymptotics for large and small distances $\sqrt{s}$ between the roots.
The first two nonzero terms of the small $s$ expansion are
\begin{equation}
\ti{\rho}_2=\frac{48}{\pi^4 (2 + |v_1 + v_2|^2)^5} s^2 + 
\frac {8 (|v_1^2 - v_2^2|^2 - 8 |v_1 - v_2|^2)}
{\pi^4 (2 + |v_1 + v_2|^2)^6} s^3 + {\cal O}(s^4).
\label{eq:smallasy}
\end{equation}
For large $s$ asymptotics in the leading term, as we expect, the two
velocities are uncorrelated, while we give also the next term of an expansion
in powers of $e^{-s}$
\begin{eqnarray}
\ti{\rho}_2 &=& \frac{4}{\pi^4}\Biggl[
\frac{1 + 6 e^{-s}}{(1+|v_1|^2)^3(1+|v_2|^2)^3}\\
&+& \frac{s^2(|v_1|^2-2)(|v_2|^2-2)
+ 12 s (|v_1|^2 + |v_2|^2-1) - 9 s |v_1+v_2|^2 - 9|v_1-v_2|^2}
{(1+|v_1|^2)^4(1+|v_2|^2)^4}e^{-s}\nonumber \\
&+& {\cal O}\left(e^{-2s}\right)\Biggr]\nonumber
\end{eqnarray}
Note that the small $s$ expansion (\ref{eq:smallasy}) should be understood
strictly pointwise, while it is not termwise integrable with respect
to velocities $v_1$ and $v_2$ as should be the case for the entire 
2-point parametric correlation function $\ti{\rho}_2$ (\ref{eq:pcf}).
It is useful to give also the velocity-moments (\ref{eq:velmom}) which in
this case depend only on the distance between zeros $\sqrt{s}$ and probably
still contain a lot of information about parametric 2-point statistics
\begin{eqnarray}
\ave{v_1 v_2^*} &=& -\sigma^2 e^s (e^s - 1)(e^{-s}-1+s)/\omega,
\label{eq:vm}\\
\ave{|v_1|^2 |v_2|^2} &=& \sigma^4 (e^s + 1)(e^s - 1)^2/\omega,\\
\ave{v_1^2 {v^*_2}^2} &=& 2\sigma^4 (e^s - 1)^3/\omega,\\
\omega &=& e^s (e^s - 1 - s)^2 + e^{2s}(e^{-s}-1+s)^2.
\end{eqnarray}

\subsection{Real coefficients}

In this subsection we discuss the case of parametric statistics of
the Gaussian random polynomials with {\em real coefficients}, i.e. $a_n$ and 
$\partial_\lambda a_n$ are {\em real Gaussian random variables} with prescribed
covariances
$$\ave{a_n a_m},\quad 
  \ave{a_n \partial_\lambda a_m},\quad
  \ave{\partial_\lambda a_n \partial_\lambda a_m}.$$
Fixing the two k-tuples of complex numbers $\ve{z}$ and $\ve{v}$ we define
$6k$ real random variables 
$\re f_j,\im f_j,\re f^{\prime}_j,\im f^{\prime}_j,\re\ti{f}_j,\im\ti{f}_j$,
or equivalently, a vector of $3\times 2k$ variables
$\ve{\xi} = (\ve{f},\ve{f}^\prime,\ti{\ve{f}})$ where
$\ve{f} = (f_1,f_1^*,\ldots,f_k,f_k^*),
\ve{f}^\prime = (f^\prime_1,{f^\prime_1}^*,\ldots,f^\prime_k,{f^\prime_k}^*),
\ti{\ve{f}} = (\ti{f}_1,\ti{f}^*_1,\ldots,\ti{f}_k,\ti{f}^*_k)$
The joint distribution of $\ve{\xi}$ is now again a (real) Gaussian with 
various blocks and their derivations of covariance matrices which are now
$2k\times 2k$ matrices and are defined by the same formulae 
(\ref{eq:defM},\ref{eq:defABC},\ref{eq:defDEF},\ref{eq:defDEF2},\ref{eq:defL},
\ref{eq:defGHK}). Using straightforward approach which follows the previous 
subsection and the derivation of nonparametric statistics for real coefficients
\cite{P96} one derives the general formula for the parametric
$k-$point correlation function of zeros of Gaussian random polynomials with
real coefficients
\begin{equation}
\ti{\rho}_k(\ve{z},\ve{v}) = \frac{1}{(2\pi)^{2k}}
\sqrt{\frac{\det\ti{\ma{L}}}{\det\ma{M}}}
\,\sper\pmatrix{
\ti{L}_{1\,1}&\ti{L}_{1\,1}&\ldots &\ti{L}_{1\,2k}&\ti{L}_{1\,2k}\cr
\ti{L}_{1\,1}&\ti{L}_{1\,1}&\ldots &\ti{L}_{1\,2k}&\ti{L}_{1\,2k}\cr
\vdots &\vdots &\ddots &\vdots &\vdots\cr
\ti{L}_{2k\,1}&\ti{L}_{2k\,1}&\ldots &\ti{L}_{2k\,2k}&\ti{L}_{2k\,2k}\cr
\ti{L}_{2k\,1}&\ti{L}_{2k\,1}&\ldots &\ti{L}_{2k\,2k}&\ti{L}_{2k\,2k}\cr}
\label{eq:realcoef}
\end{equation}
which reduces to the corresponding 
nonparametric $k-$point correlation function \cite{P96}
\begin{equation}
\rho_k(\ve{z}) = \frac{\sper\ma{G}}{(2\pi)^k\sqrt{\det\ma{A}}}
\end{equation}
after velocities are integrated out.
{\em Semi-permanent} of a square $2 m\times 2 m$ matrix
(introduced in \cite{P96}) is a homogeneous 
polynomial of order $m$ of the matrix elements 
\begin{equation}
\sper\ma{S} =
\sum_{j_1 < \ldots < j_m \atop l_1 < \ldots < l_m}^{j_n \neq l_{n^\prime}}
\sum\limits_{p} \prod\limits_{r=1}^m S_{j_r + m,l_{p_r}}. 
\label{eq:sp}
\end{equation}
where $p$ runs over all $m!$ permutations of $m$ indices $\{1\ldots m\}$ and
addition of indices $j_r+m$ should be understood modulo $2m$.
Note that 
$\sper\left(\pmatrix{1&1\cr 1&1}\otimes\ti{\ma{L}}\right)
\neq\sper\left(\ti{\ma{L}}\otimes\pmatrix{1&1\cr 1&1}\right)$,
whereas 
$\per\left(\pmatrix{1&1\cr 1&1}\otimes\ti{\ma{L}}\right)
=\per\left(\ti{\ma{L}}\otimes\pmatrix{1&1\cr 1&1}\right)$,
so the $4k\times 4k$ matrix on the RHS of eq. (\ref{eq:realcoef}) cannot be
written as $\pmatrix{\ti{\ma{L}}&\ti{\ma{L}}\cr\ti{\ma{L}}&\ti{\ma{L}}}$
like in the complex case (\ref{eq:parstc}).

Since the case of real coefficients is much more complicated than the case
of complex coefficients we consider only the final and most important 
specialization, that is of harmonic random polynomials with real 
coefficients in the `thermodynamic' limit $N\rightarrow\infty$, 
$\ave{a_n a_m} = \delta_{nm}/n!,\ave{a_n\partial_\lambda a_m} =0,
\ave{\partial_\lambda a_n \partial_\lambda a_m} = 
\sigma^2\delta_{nm}/n!, g(s) = \exp(s), \bar{g}(s) = \sigma^2\exp(s)$.
Then it is easy to see (like in the nonparametric case \cite{P96}) that the
two cases of real and complex coefficients are different only when some of
the zeros $z_j$ is close to the real (symmetry) axis, whereas in the opposite 
case, where all $z_j$ go away from the real axis, the parametric statistics for
the case of real coefficients converge to parametric statistics for the
complex coefficients. 
Explicit formulae for $1$ and $2$ point parametric correlation functions for
this case are too lengthy to write out, we shall only give an asymptotic 
expansion for $1-$point function close to the real axis (small $y = \im z$)
\begin{equation}
\ti{\rho}_1(x+iy,u+iw) = \frac{48\sigma y^2}{\pi^2 (\sigma^2 + 2 u^2)^{3/2}} 
- \frac{8\sigma(u^4 + 3 u^2 w^2)y^4}
{\pi^2(\sigma^2 + 2 u^2)^{7/2}} + {\cal O}(y^6)
\label{eq:smalls}
\end{equation} 
It is interesting to calculate the `root mean square' (RMS) velocity which now 
depends on the distance from the real axis $y$
\begin{equation}
\ave{|v|^2} = \sigma^2 \frac{e^{4 y^2} - 1}{e^{4 y^2} - 1 - 4 y^2}
\end{equation}
The fact that RMS velocity has a singularity around $s=0$, 
$\ave{|v|^2}\propto s^{-2}$, is compatible with finding that the small
$s$ expansion of the parametric density (\ref{eq:smalls}) 
is not termwise integrable with respect to the velocity.

\section{Numerical results}

The predictions of the theory of random polynomials have been verified on the
parametric statistics of zeros of Husimi representation of eigenfunctions
of a generic chaotic autonomous Hamiltonian system with two freedoms which 
depends on an external parameter. We have chosen a recently introduced
semi-separable oscillator for which an extremely efficient quantization scheme
exists based on a construction of an exact quantum Poincare mapping 
\cite{P95a,P95}. The Hamiltonian of the system is the following
\begin{equation}
H = -\half\hbar^2\partial_x^2 - \half\hbar^2\partial_y^2 + 
\half (x - a\,\sgn{y})^2
\end{equation}
where the configuration space is a 
strip $(x,y)\in (-\infty,\infty)\times [b_\downarrow,b_\uparrow]$ 
with hard walls and therefore Dirichlet boundary conditions for a 
wavefunction at $y=b_\downarrow,b_\uparrow$. The system is integrable for 
$a=0$ while for $a>0$ the potential becomes discontinuous and the system
becomes increasingly chaotic with increasing $a$. It has been shown in
\cite{P95} that the system is fully chaotic (with possible islands of
stability being negligibly small) for the following values of parameters:
$a=0.25,b_\uparrow=4,b_\downarrow=-11,E=0.5$. For these values and for
$\hbar=0.0003$ we have calculated a stretch of $16$ consecutive eigenstates
with sequential quantum number (according to Thomas-Fermi rule) 
$\approx 17\,684\,000$. Then we have calculated a second set of eigenstates
according to variation of parameters $a=0.25,b_\uparrow=4+\lambda,
b_\downarrow=-11+\lambda$ with $\lambda = 5\cdot 10^{-7}$. For each of the
eigenstates of the two sets we have calculated the Husimi representation on
the surface of section $(x,p_x)$ \cite{P95}, its zeros (only inside classically
allowed (chaotic) region of phase space) and calculated the variations of 
zeros w.r.t. small variation of the parameter from $\lambda=0$ to $\lambda=
5\cdot 10^{-7}$. Since the variation of the external parameter
$\delta\lambda=5\cdot 10^{-7}$ was small enough, the identification of
corresponding zeros has always been possible and the velocities were 
numerically well defined. Assuming the conjecture that the parametric 
statistics of zeros of Husimi representation of a chaotic eigenstate should
be described by the theory of random polynomials, only two 
parameters remain which define the scales, namely the average density of
zeros $\rho_1$ (which should be constant inside the classically allowed chaotic 
region of surface of section) and the root mean square velocity $\sigma=
\sqrt{\ave{|v|^2}}$. The density or the number of inside zeros is roughly
constant for our 16 states while RMS velocity $\sigma$, which is determined
by the sensitivity of a given eigenstate with respect to the variation of
external parameter $\lambda$, might exhibit substantial fluctuations. 
Therefore, if we want to merge numerical data of all eigenstates together
in order to improve statistics, we should first rescale the data for 
each individual eigenstate to a unit RMS velocity. 

In figure 1 we plot such numerical
velocity distribution for our stretch of 16 eigenstates which contains
velocities for $\approx 16\times 1800 = 28800$ zeros (only the zeros whose
distance from the boundary of classically allowed region of surface of 
section was larger than few mean spacings were taken into account).  
It is evident that the agreement with the prediction of theory
of random polynomials (\ref{eq:veldis}) is statistically significant.

In order to compare significantly the numerical estimate of the parametric 
$2-$point 
(or higher) statistics with the prediction of the theory of random polynomials
one needs a data of a higher statistical quality than those of our
numerical experiment. This may be easier in some explicit one-dimensional
quantum maps and we leave it as a challenge for a future work.

\section{Zero-flow as an integrable dynamical system}

In this section we show explicitly that the flow of zeros of a polynomial
with respect to some external parameter $\lambda$ interpreted as a 
fictitious `time' can be written in terms of a closed set of 
`equations of motion'. We argue that the underlying dynamical system
is completely integrable and construct explicitly the complete set of
integrals of motion as functions of zeros and velocities. 
This construction should be considered as a polynomial analog of
Dyson-Pechukas-Yukawa \cite{P83,Y86} level dynamics of matrices.

Let us consider a polynomial of order $N$ whose dependence on the
external parameter $\lambda$ is linear
\begin{eqnarray}
f(z,\lambda) &=& \prod_{j=1}^N (z - z_j(\lambda)) = 
f_0(z) + \lambda f_1(z), \label{eq:deff2}\\
f_0(z) &=& \sum_{n=0}^N a^0_n z^n,\quad 
f_1(z) = \sum_{n=0}^N a^1_n z^n.\nonumber
\end{eqnarray}
where without essential loss of generality we have assumed that
\begin{equation}
a^0_N = 1,\quad a^1_N = 0.
\end{equation}
Writing the velocities 
$v_j = (d/d\lambda)z_j = -f_1(z_j)/\prod_{l\neq j}(z_j-z_l)$,
differentiating with respect to $\lambda$ again and using Lagrange
interpolation formula for $f_1(z_j)$ and $f^\prime_1(z_j)$
one arrives to the simple closed set of equations for $2N$ complex
(or $4N$ real) dynamical variables, zeros $z_j$ and
velocities $v_j$, $j=1,\ldots,N$
\begin{eqnarray}
\frac{d}{d\lambda} z_j &=& v_j,\nonumber \\
\frac{d}{d\lambda} v_j &=& 2 \sum_{k\neq j}\frac{v_j v_k}{z_j - z_k}.
\label{eq:eqm}
\end{eqnarray}
From the construction it is obvious that the underlying dynamical system 
should be completely integrable. Writing the polynomial 
\begin{equation}
f(z,\lambda) = \prod_{j=1}^N (z - z_j) = 
\sum_{n=0}^N (-1)^n c_n(z_1,\ldots,z_N) z^{N-n}
\label{eq:defc}
\end{equation}
in terms of fully symmetric homogeneous functions 
\begin{equation}
c_n(z_1,\ldots,z_N) = \sum_{{\cal A}\subset\{1,\ldots,N\}}^{|{\cal A}|=n}
\prod_{j\in {\cal A}} z_j
\end{equation} 
one easily sees that their derivatives
\begin{equation}
I_n(z_1,\ldots,z_N,v_1,\ldots,v_N) = \frac{d}{d\lambda}c_n(z_1,\ldots,z_N) = 
\sum_{j=1}^N v_j\partial_{z_j}c_n 
\end{equation}
are just equal to the coefficients of the perturbation polynomial (see eqs.
(\ref{eq:deff2},\ref{eq:defc})
\begin{equation}
I_n = (-1)^n a^1_{N-n}
\end{equation}
and therefore independent nontrivial (complex) constants of motion, for 
$n=1,\ldots N$.
It seems more difficult to give a simple Lagrangian or Hamiltonian formulation
of the equations (\ref{eq:eqm}). Since the system is completely integrable
its Lagrange function is not unique. However, it is easy to show that eqs. 
(\ref{eq:eqm}) 
are completely equivalent to Euler-Lagrange equations for any Lagrange function
$L(z_1,\ldots,z_N,v_1,\ldots,z_N)$ which can be written in terms of
a {\em symmetric (real)} and {\em nondegenerate} holomorphic function of constants 
of motion and their complex conjugates
\begin{eqnarray}
L&=&F(I_1,\ldots,I_N,I^*_1,\ldots,I^*_N),\\
F(I_1,\ldots,I_N,I^*_1,\ldots,I^*_N) &=& F(I^*_1,\ldots,I^*_N,I_1,\ldots,I_N),
\nonumber\\
\det\frac{\partial^2 F}{\partial I_j\partial I^*_k} &\neq& 0. \nonumber
\end{eqnarray}

\section{Conclusions}

We have defined a new type of local 
parametric statistics of quantum eigenstates,
which describes the statistical distributions of zeros of Husimi functions 
and the velocities --- the derivatives of zeros with respect to an external 
parameter. It has been conjectured that apart from the scaling parameters 
(density of zeros and mean square velocity) the parametric statistics of
zeros of a Husimi representation of an eigestate of a classically chaotic 
system should behave universally and should be well described by the 
ensembles of Gaussian random polynomials to the same extent as the statistics 
of energy spectra, eigenvector components, or matrix elements are 
described by the ensembles of Gaussian random matrices. The parametric 
statistics of roots of Gaussian random polynomials (or zeros of
Gaussian random holomorphic functions) have been solved
analytically and its applicability to simple chaotic Hamiltonian systems
has been demonstrated numerically.

In case where the dependence upon external parameter is linear
the flow of zeros of an analytic (Bargmann) function has been formulated in 
terms of fully integrable dynamical system, where the 
external parameter and the zeros are playing the roles of fictitious time 
and mutually repelling quasi particles, respectively.
The analytical results on local parametric statistics presented in the
first part of this paper are nothing else but `equal-time' 
correlation functions of such a dynamical system. The fact that the
velocity distribution (\ref{eq:veldis}) is a generalized Lorentzian and
not Maxwellian is due to nonseparability of kinetic and potential energy.
For example, the $2-$body case ($N=2$) in the center of mass frame
$v=v_1=-v_2=2\dot{z},z=z_2-z_1$ can be described by the Hamiltonian
$$H=\half |\dot{z}|^2|z|^2 = \half |p|^2/|z|^2,$$ 
where $p = |z|^2 \dot{z}$ is a (complex) momentum conjugated to the 
(complex) coordinate $z$. 

The generalization to the calculation of the most general `different-time' 
correlation function --- {\em non-local} parametric $k-$point correlation 
function 
$$\ti{\rho}(\ve{z}^\prime,\ve{v}^\prime,\ve{\lambda})
=\ave{\prod\limits_{j=1}^k \delta(z^\prime_j - z_j(\lambda_j))
\delta(v^\prime_j-(d/d\lambda)z_j(\lambda_j))}
$$
is straightforward by putting 
$$f_j = f(z_j,\lambda_j),\quad
f^\prime_j = \partial_z f(z_j,\lambda_j),\quad
\ti{f}_j = (d/d\lambda) f(z_j,\lambda_j)$$
instead of (\ref{eq:deff},\ref{eq:defdf},\ref{eq:deftif}) and following the
formalism of section 2.

\section*{Acknowledgements}

The hospitality of the Institut Henri Poincar\' e, Paris, where the
major part of this work has been done, and the
financial support of C.I.E.S. (France) and the Ministry of Science and
Technology of the Republic of Slovenia are gratefully acknowledged.

\section*{Figure captions}

\noindent {\bf Figure 1:}
The numerical velocity distribution of the chaotic semiseparable oscillator 
(full curve) (see text for details) as compared to the theoretical velocity
distribution (\ref{eq:veldis}) based on the theory of random polynomials 
(dashed curve). The agreement is highly statistically significant.
\end{document}